\documentclass[fleqn,twoside,twocolumn,nofootinbib]{revtex4} 
\usepackage{ujp} 
\begin{document}
\title[EFFECT OF WEAK MAGNETIC FIELD ($\sim $\,300 Gs)]
{EFFECT OF WEAK MAGNETIC FIELD (\boldmath$\sim $\,300 Gs)\\ ON THE
INTENSITY OF TERAHERTZ EMISSION\\ OF HOT ELECTRONS IN
\boldmath$n$-Ge AT HELIUM
TEMPERATURES}%
\author{V.M. Bondar}
\affiliation{Institute of Physics, Nat. Acad. of Sci. of Ukraine}
\address{46, Nauky Ave., Kyiv 03680, Ukraine}
\email{ptomchuk@iop.kiev.ua}
\author{P.M. Tomchuk}
\affiliation{Institute of Physics, Nat. Acad. of Sci. of Ukraine}
\address{46, Nauky Ave., Kyiv 03680, Ukraine}
\author{G.A. Shepel'skii}
\affiliation{Lashkaryov Institute of Semiconductor Physics, Nat. Acad. of Sci. of Ukraine}
\address{41, Nauky Ave., Kyiv 03680, Ukraine}
 \udk{???}  \razd{\secix}

\setcounter{page}{1020}%
\maketitle

\begin{abstract}
 Experimental results of studying the effect of a weak magnetic
field ($\sim $300 Gs) on the intensity of the terahertz emission
($\lambda \approx $100 $\mu $m) of hot electrons in $n$-Ge
(crystallographic orientation $\langle 1,0,0 \rangle)$ at helium
temperatures ($T\sim $5 K) are presented and discussed. It is shown
that the strong influence of this field (decrease of the emission
intensity by 500$\div $1000{\%}) is related to a decrease of the
carrier concentration at weak electric fields and the appearance of
the magnetoresistance at stronger fields. The longitudinal
magnetoresistance becomes significant due to the anisotropy of the
energy dispersion law of electrons and a strong deformation of the
electron velocity distribution function by the electric field (which
is beyond the framework of the diffusion approximation).
\end{abstract}

\section{Introduction}

Recently, the peculiarities of mechanisms of generation and
absorption of terahertz light attract still more attention of
investigators [1]. In [2--4], we studied the angle dependences of
the terahertz emission of hot electrons in $n$-Ge. This
semiconductor has a cubic symmetry, but, in the case where the
electric field is directed non-symmetrically with respect to valleys
(minima in the conduction band), electrons in different valleys can
have different temperatures. This results in the symmetry violation
and, consequently, the appearance of the polarization dependence of
the hot electron emission. We studied the relation between the
polarization dependences and anisotropic scattering mechanisms [4,
5] characteristic of many-valley semiconductors. It was a surprise
to discover that, under certain conditions (low temperatures, strong
electric fields), the polarization dependences of the hot electron
emission appear in the case where the electric field is oriented
along the $\langle 1,0,0\rangle$ direction, i.e. symmetrically with
respect to valleys. It was established that, in this case, the
appearance of the polarization dependences is related to the
symmetry violation of the even part of the electron velocity
distribution function under the action of an electric field. In
other words, this effect can be explained going beyond the bounds of
the traditional so-called diffusion approximation. As is known (see,
e.g., [6]), this approximation is based upon the smallness of the
ratio of the drift velocity of an electron to its mean thermal
velocity. At low temperatures and strong electric fields, the
diffusion approximation appeared invalid. Another surprise was
aroused by an extremely high sensitivity of the hot electron
emission intensity to weak magnetic fields ($H\sim $~300 Gs) at low
temperatures ($T\sim $~5 K). The emission intensity can fall by an
order of magnitude due to the application of the magnetic field.
This work is devoted to the study and explanation of this
phenomenon.

\section{Experimental Part}

All measurements were performed on the set-up described in [4]
supplemented with an attachment allowing one to subject an emitting
sample to the action of the magnetic field of the required direction
and magnitude -- from zero to the maximum value. This field was
created with the use of a permanent magnet with the corresponding
devices used to regulate the field intensity. The arrangement of the
units is schematically shown in Fig. 1.

The $n$-Ge samples were cut off in the crystallographic directions
$\langle 1,1,1\rangle$ or $\langle 1,0,0\rangle$, had a standard
size of 7$\times $1$\times $1 mm$^{3}$, and were treated using the
standard technique [4]. The electric field was created by pulses
with a duration of 0.8 $\mu $s and a repetition rate of 6 Hz. After
that, the signal of a semiconductor detector was amplified,
integrated, and converted to the direct-current voltage proportional
to the intensity of the hot electron emission of the sample in the
region $\lambda \approx $100 $\mu $m. The ohmicity of the contacts
to $n$-Ge was provided using St alloy with a 5-{\%} fraction of Sb.

\begin{figure}
\includegraphics[width=7cm]{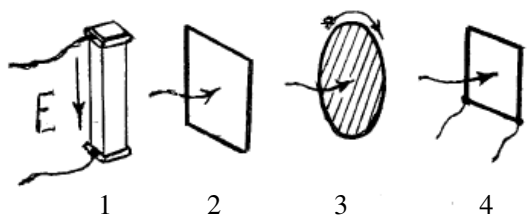}
\vskip-3mm\caption{Diagram of the experiment. {\it 1} -- $n$-Ge
sample; {\it 2} -- filter limiting high frequencies; {\it 3} --
rotating polarizer; {\it 4} -- Ge(Ga) receiver  }
\end{figure}

\begin{figure}
\includegraphics[width=6cm]{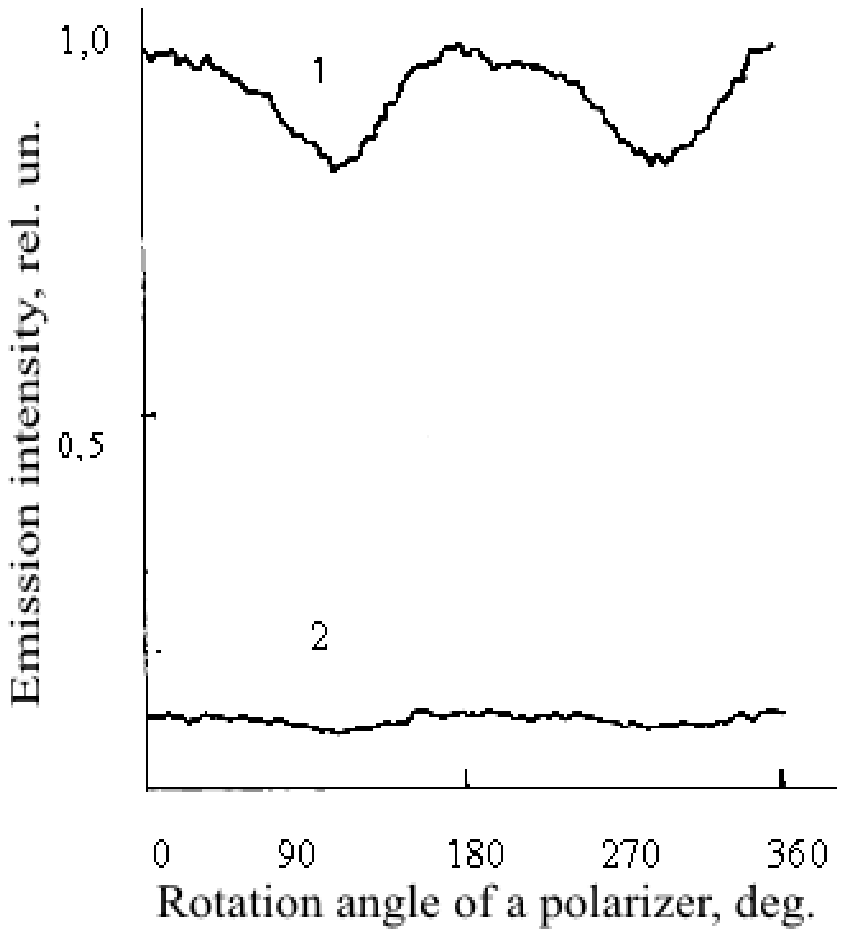}\\
{\bf a}\\ \vskip2mm
\includegraphics[width=6cm]{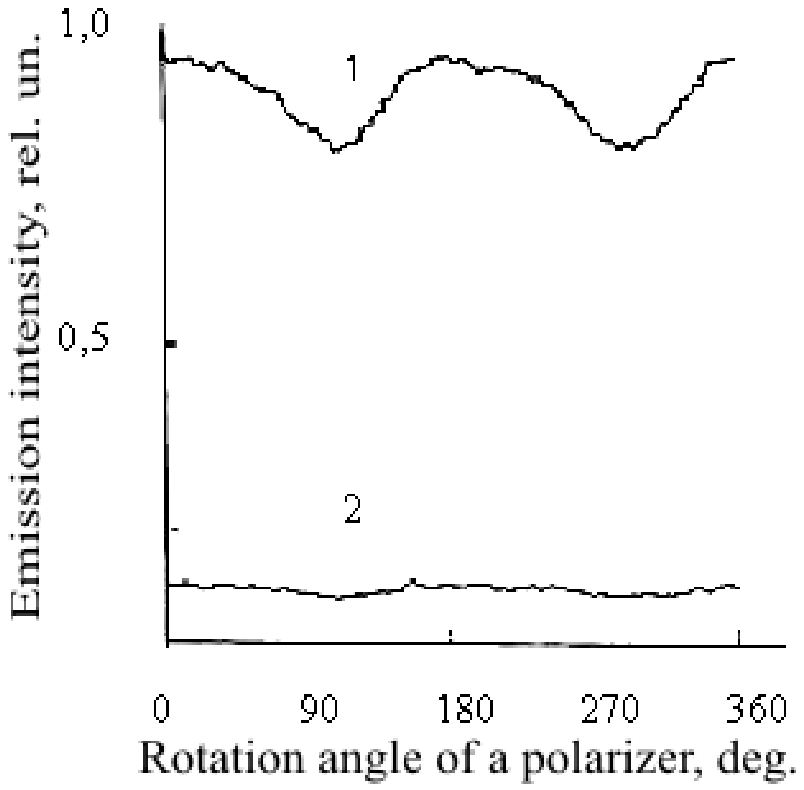}\\
{\bf b}\\ \vskip2mm
\includegraphics[width=6cm]{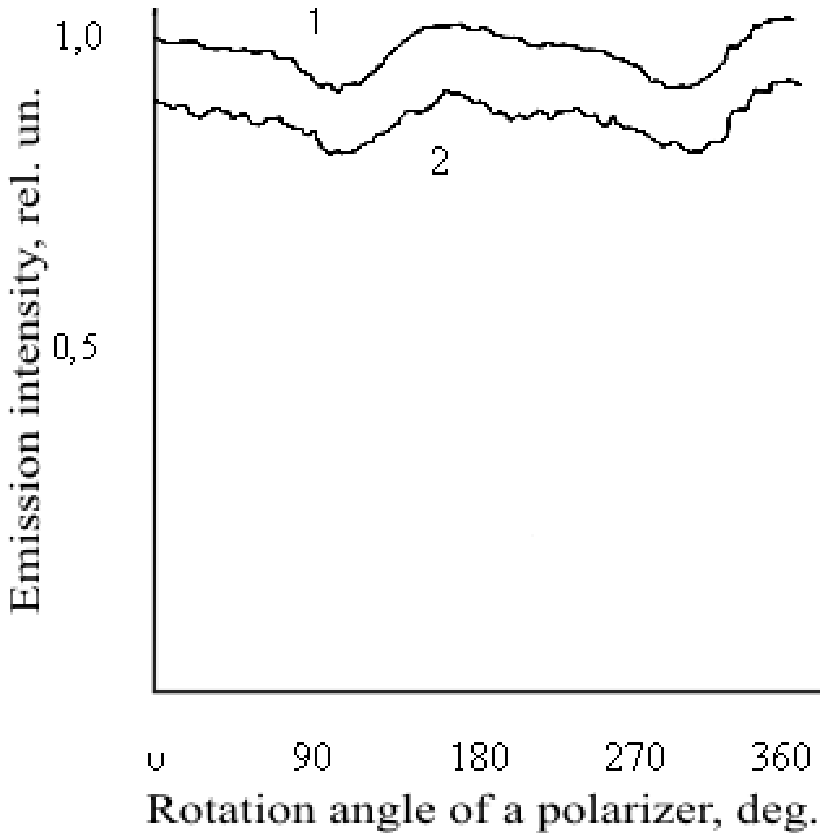}\\
{\bf c}\\ \vskip-3mm\caption{Receiver signal: {\it 1} -- without
magnetic field {\it 2} -- with weak magnetic field; heating electric
fields: 15 V/cm ({\it a}), 25 V/cm ({\it b}),
 200 V/cm ({\it c})  }
\end{figure}

\begin{table*}[!]
\noindent\caption{ }\vskip3mm\tabcolsep12.5pt

\noindent{\footnotesize\begin{tabular}{c c c c c c c c c}
 \hline \multicolumn{1}{c}
{\rule{0pt}{9pt}$V$} &\multicolumn{1}{|c} {2}& {3}& {4}& {5}& {9}&
{15}&{30}& {45}\\%
\hline%
\multicolumn{1}{c}{$n$}&\multicolumn{1}{|c}{3.21$\times
$10$^{9}$}&7$\times $10$^{9}$&2.8$\times $10$^{12}$&1.41$\times
$10$^{13}$& 9.3$\times $10$^{13}$& 1.7$\times
$10$^{14}$&6.7$\times $10$^{14}$&2.3$\times $10$^{14}$ \\
\hline
\end{tabular}}
\end{table*}

\section{Experimental Results and Their Discussion}

Figure 2,{\it a--c} presents the experimental results of studying
the effect of a weak magnetic field on the intensity of the hot
electron emission in $n$-Ge. One can see that, at small electric
fields, the magnetic field reduces the emission amplitude by almost
an order of magnitude. With increase in the electric field, the effect
of the magnetic field on the emission intensity becomes considerably
weaker. Such changes in the emission intensity under the action of
the weak magnetic field cannot be explained by its influence on the
dispersion law or scattering mechanisms. Simple estimates
demonstrate that, during the mean free time between collisions of a
carrier with scattering centers, its trajectory changes
insignificantly. In this connection, it was necessary to search for
other reasons of such a strong effect of the weak magnetic field on
the terahertz emission intensity.

For this purpose, we studied the electrophysical characteristics of
the $n$-Ge samples investigating their emission at helium
temperatures. We took volt-ampere characteristics and carried out
Hall measurements in order to determine the carrier concentration
starting from low voltages, at which not all donors are ionized [7].
The measuring results are given in Fig. 3 and Table. As follows from
the behavior of the volt-ampere characteristic of the sample in the
absence and in the presence of the magnetic field (we recall that
the magnetic field is weak, close to 300 Gs), the resistance of the
sample in such a field grows almost by an order of magnitude at an
electric field of $\sim $5 V. With increase in the electric field,
the resistance of the sample grows much more slowly: only by tens of
percent. One can trace a direct connection between the increase of
the sample resistance and the fall of the emission intensity.

\begin{figure}
\includegraphics[width=6cm]{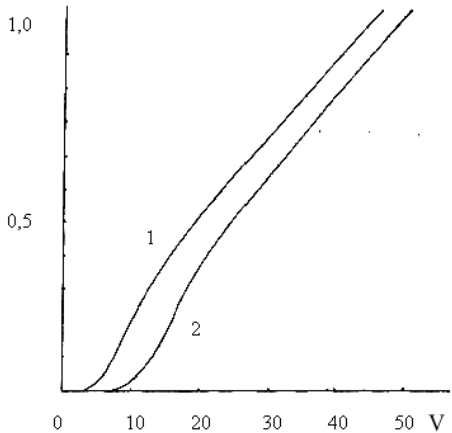}
\vskip-3mm\caption{Volt-ampere characteristic of the sample: {\it 1}
-- without magnetic field {\it 2} -- with weak magnetic field ($\sim
$300 Gs) }
\end{figure}

Thus, the first part of our task aimed at clarifying the reasons of
a decrease of the terahertz emission intensity under the action of
the weak magnetic field is solved, though far from exhaustively.
Now, we need to explain the large growth of the sample resistance
under the action of such a weak magnetic field at helium
temperatures. It turned out that similar phenomena have been already
studied and were explained by the hopping conduction in the impurity
band. According to the existing conceptions (see, e.g., [8]),
the main contribution into the conduction at low temperatures (at
which the majority of electrons are localized at impurities) is made
by the hopping mechanism. The effect of the weak magnetic field on
this mechanism is explained by its influence on the ``tails'' of the
wave function of electrons localized at donors. The overlapping of
these ``tails'' determines the probability of hops of electrons to
vacancies.

It is worth noting that the thermal introduction of carriers into the conduction
band is ineffective at low temperatures, that is why the hopping conduction
over vacancies is provided by the compensation effect. This compensation is
present in practically any material. Assuming that the impurity breakdown
starts according to the Zener mechanism, it is easy to understand the effect
of the magnetic field on the free carrier concentration in the conduction
band on the stage where all donors are non-ionized. In turn, this
explains the influence of the weak magnetic field on the hot electron
emission at low temperatures.

In addition, the ``attachment'' of the
carriers already introduced into the conduction band to neutral donors,
the inverse process, and the dependence of both processes on the magnetic
field are possible.

All the above-said concerns a decrease of the emission under the
action of a magnetic field at low electric fields 10$\div $15 V/cm
(Fig. 2,$a)$.

At strong electric fields, this decrease is much smaller (Fig.
2,{\it c}) and amounts to $\sim $10{\%} of the initial value. Such a
behavior of the observed phenomenon can be explained by a
deformation of the velocity distribution function in the case where
the electric field is oriented along the $\langle 1,0,0\rangle$
direction and develops into the heating one. In this case, the
diffusion approximation can explain fine characteristics of the
discussed phenomena not always, and one should use a more accurate
distribution function.

At strong electric fields (at which the concentration of carriers in the
conduction band does not change anymore), the effect of a magnetic field
on the hot electron emission is related to the appearance of the
longitudinal magnetoresistance. The latter is connected with a decrease in
the electron heating and therefore a fall of the emission. The mechanism
of the formation of the longitudinal magnetoresistance in many-valley
semiconductors is considered in the following section.

\section{Longitudinal Magnetoresistance}

For today, the general theory of galvanomagnetic phenomena in many-valley
semiconductors with regard for the anisotropy of the dispersion law of
carriers and mechanisms of their scattering is well developed and solidly
substantiated (see, e.g., [9]). However, the general formulas of this
theory are rather cumbersome. The situation becomes still more complicated
when trying to allow for the possibility of the electron heating by the
electric field.

The purpose of this work is narrower -- to explain the reasons for
the appearance of the longitudinal magnetoresistance and estimate
its magnitude in the case where arbitrary electric and weak magnetic
fields are oriented along the direction symmetric with respect to
valleys ($\langle1,0,0\rangle$ in $n$-Ge). That is why we can employ
a rougher but simpler model. The essence of this approximation is to
characterize the hot electrons by a velocity-shifted Maxwellian
distribution function [10] or (in the case of degeneracy) by the
Fermi [11] function with the effective electron temperature. In
many-valley semiconductors, such a function is to be introduced for
electrons of each valley. In the general case where there exists the
possibility of a degeneracy of the electron gas in the $\alpha $-th
ellipsoid, we can write [11]
\begin{equation}
\label{eq1} f_\alpha =\left\{ {1+\exp \left( {+\frac{\varepsilon (
\boldsymbol{\upsilon })- \mathbf {p} \mathbf {u}^{(\alpha )}-\mu
^{(\alpha )}}{kT^{(\alpha )}}} \right)} \right\}^{-1},
\end{equation}
where $\mathbf {\upsilon }$ is the electron velocity, $\varepsilon
(\mathbf {\upsilon })$ and $\mathbf {p}$ are its energy and
momentum, respectively, $T^{(\alpha )}$ denotes the effective
electron temperature, $\mu ^{(\alpha )}$ is the chemical potential,
and $\mathbf {u}^{(\alpha )}$ is the drift velocity. The quantities
$\mu ^{(\alpha )}$, $T^{(\alpha )}$, and $\mathbf {u}^{(\alpha )}$
must be determined from the equations for the concentration, energy,
and momentum balance, respectively. In what follows, we will
consider the case where the electric field $\mathbf {E}$ and the
magnetic field $\mathbf {H}$ are oriented along the direction
symmetric with respect to valleys ($\langle1,0,0\rangle$ for
$n$-Ge), so the parameters $\mu ^{(\alpha )}$ and $T^{(\alpha )}$
will be the same for all valleys. As concerns the drift velocity
$\mathbf {u}^{(\alpha )}$ in the given symmetric case, it will have
the same absolute value but different directions for different
valleys. That is why we do not present explicitly the balance
equations for the electron concentrations in valleys and their
energies and restrict ourselves to the momentum balance equation. In
the principal axes of the $\alpha $-th mass ellipsoid, in which the
energy dispersion law has the standard form
\begin{equation}
\label{eq2} \varepsilon ( \boldsymbol{\upsilon })=\frac{P_\perp ^2
}{2m_\perp }+\frac{P_{\parallel}^2 }{2m_{\parallel } },
\end{equation}
the momentum balance equation is
\begin{equation}
\label{eq3} e\left\{ {E_i +\frac{1}{c}\left[ { \mathbf {u}^{(\alpha
)}\,\times  \mathbf {H}} \right]_i =\frac{m_ \perp }{\tau _ \perp
}\,\,u_i^{(\alpha )} ,(i=x,y)} \right\},
\end{equation}
\begin{equation}
\label{eq4} e\left\{ {E_z +\frac{1}{c}\left[ { \mathbf {u}^{(\alpha
)}\,\times  \mathbf{H}} \right]_{z} =\frac{m_{\parallel} }{\tau
_{\parallel} }\,\,u_z^{(\alpha )} } \right\},
\end{equation}
where $e$ is the electron charge, $c$ is the light velocity, while
$\tau _{\parallel} $ and $\tau _\perp $ are the longitudinal and
transverse relaxation times, respectively.

Due to the weakness of the magnetic field, we can solve Eqs. (\ref{eq3}) and (\ref{eq4}) using
perturbation theory with respect to the parameter $H$. In the zero-order
approximation (i.e. at $H=0)$, Eqs. (\ref{eq3}) and (\ref{eq4}) yield
\begin{equation}
\label{eq5} \left\{ {u_i^{(\alpha )} } \right\}_0 =\frac{e\tau _
\perp}{m_ \perp }\,E_i ,\;\;\,i=x,y,
\end{equation}
\begin{equation}
\label{eq6} \left\{ {u_z^{(\alpha )} } \right\}_0 =\frac{e\tau
_{\parallel}}{m_{\parallel} }\,E_z,
\end{equation}
\begin{equation}
\label{eq7}
 \mathbf {u}_0^{(\alpha )} =\frac{e\tau _ \perp }{m_ \perp } \mathbf {E}+\left(
{\frac{e\tau _{\parallel} }{m_{\parallel} }-\frac{e\tau _ \perp }{m_
\perp }} \right)\left( { \mathbf {i}_\alpha  \mathbf {E}} \right)
\mathbf {i}_\alpha ,
\end{equation}
where $\mathbf {i}_\alpha $ is the unit vector that specifies the
orientation of the $\alpha$-th ellipsoid (valley). From
Eq.(\ref{eq7}) (or Eqs.(5)--(6)), one can see that the direction of
the drift velocity $\mathbf {u}^{(\alpha )} $ does not coincide with
that of the electric field, unless this field is directed along the
principal axis of the mass ellipsoid. As a result, the term
[$\mathbf {u}^{(\alpha )} \times \mathbf {H}$] will be non-zero in
spite of the fact that $\mathbf {H}\parallel \mathbf {E}$. This is
the reason for the appearance of the longitudinal magnetoresistance
in $n$-Ge.

Then, one can develop the perturbation theory with respect to $H$,
i.e. substitute the approximate value $\mathbf {u}_0^{(\alpha )}
$in the term [$\mathbf {u}^{(\alpha )}\times \mathbf {H}$]
neglected when obtaining Eq.(\ref{eq7}), which yields $\mathbf
{u}_1^{(\alpha )} $ and so on. As a result, we obtain the series
with respect to $H$ for the drift velocity of electrons of the
$\alpha$-th valley (see Appendix):
\begin{equation}
\label{eq8}
 \mathbf {u}^{(\alpha )}=
\quad
 \mathbf {u}_0 ^{(\alpha )}+
 \mathbf {u}_1 ^{(\alpha )}+
 \mathbf {u}_2 ^{(\alpha )}+\ldots .
\end{equation}
The term linear in $H$ in the drift velocity (\ref{eq8}) ($\mathbf
{u}_1 ^{(\alpha )})$ determines the Hall current. Due to the
symmetry of the problem, the total Hall current in all valleys is
equal to zero. The component $\mathbf {u}_2 ^{(\alpha )}$ determines
the magnetoresistance. Its $\langle 1,0,0 \rangle$-projection and
the sum over all valleys determines the addition $\Delta J_2 $ to
the current $J_0 =J(H=0)$. As is shown in Appendix,
\begin{equation}
\label{eq9} \frac{\Delta J_2 }{J_0 }= -\frac{(e^2\,\tau _ \perp /m_
\perp \,c^2)(\tau _{\parallel} /m_{\parallel} -\tau _ \perp /m_
\perp )^2\cdot {\mathrm{H}}^2}{3(\tau _{\parallel} /m_{\parallel}
+2\tau _ \perp /m_ \perp )},
\end{equation}
where
\begin{equation}
J_0 =\frac{e^2n}{3}\left( {\frac{\tau _{\parallel} }{m_{\parallel}
}+2\frac{\tau _ \perp }{m_ \perp }} \right),
\end{equation}
and $n$ is the total electron concentration (in all valleys).

For $n$-Ge,
\begin{equation}
m_ \perp \ll  m_{\parallel}, \quad \tau _ \perp \sim \tau
_{\parallel}.
\end{equation}
In this case, formula (\ref{eq9}) acquires the simplified form
\begin{equation}
\label{eq10} \frac{\Delta J_2 }{J_0 }= - \frac{1}{6}\frac{e^2\,\tau
_ \perp}{m_ \perp ^2 c^2}{{H}}^2.
\end{equation}
Assuming for the estimate that $m_\perp \approx 0.7\times 10^{-28}$~g,
$\tau _\perp \approx 10^{-11}\,$~s, and $H\approx 300$~Oe, we obtain
from Eq.(\ref{eq10}) that $\Delta J_2^ /J_0 \approx -1/12$, which is
in good agreement with the experiment.

\section{Conclusions}

The performed studies allow us to make the following
conclusions. At low temperatures and weak electric fields, at which
the majority of electrons is localized at donor levels, the effect
of a weak magnetic field on the volt-ampere characteristics and
the emission is explained as follows. Both the hopping conduction and
the Zener breakdown mechanism are sensitive to the influence of a
magnetic field on the ``tails'' of the wave function of an electron
localized at a donor. This explains the fast decrease of the current
and the emission due to the application of a weak magnetic field. In
strong electric fields, all donors are ionized and the electron
concentration in the conduction band is constant. The effect of the
magnetic field on the volt-ampere characteristics and the emission in
strong electric fields is much weaker. In this case, such a weak
influence is explained by the longitudinal magnetoresistance. The
mechanism of the longitudinal magnetoresistance is related to the
anisotropy of the dispersion law of electrons in $n$-Ge.

\vskip3mm
 In conclusion, the authors express their gratitude to O.G.
Sarbey and S.M. Ryabchenko for the discussion of a number of
questions.

\subsubsection*{APPENDIX}

{\footnotesize With the use of Eq.(\ref{eq7}), we obtain:
\begin{equation*}
[ \mathbf {u}_0^{(\alpha )} \times  \mathbf {H}]=e(\tau _{\parallel}
/m_{\parallel} -\tau _ \perp /m_ \perp )( \mathbf {i}_\alpha \,
\mathbf {E})\left[ { \mathbf {i}_\alpha \times  \mathbf {H}}
\right].\tag{A1}
\end{equation*}
Assuming that the vector [$\mathbf {i}_\alpha\times \mathbf {H}$] is
directed along the $x$ axis and substituting Eq. (А1) into
(\ref{eq3}), one can put down
\begin{equation*}
\mathbf {u}_1^{(\alpha )} =\frac{e^2\tau _ \perp }{m_ \perp
c}\,e(\tau _{\parallel} /m_{\parallel} -\tau _ \perp /m_ \perp )(
\mathbf {i}_\alpha  \mathbf {E})\left[ { \mathbf {i}_\alpha \times
 \mathbf {H}} \right].\tag{A2}
\end{equation*}
After that, Eq. (А2) yields
\[
 [ \mathbf {u}_1^{(\alpha )} \times  \mathbf {H}]=\frac{e^2\tau _ \perp }{m_ \perp
c}(\tau _{\parallel} /m_{\parallel} -\tau _ \perp /m_ \perp )(
\mathbf {i}_\alpha  \mathbf {E})\left[ {( \mathbf {i}_\alpha \times
\, \mathbf {H})\times \mathbf {H}} \right]=
\]
\begin{equation*}
 =\frac{e^2\tau { }_ \perp }{m_ \perp c}(\tau {
}_{\parallel}\,/m_{\parallel})
 ( \mathbf {i}_\alpha  \mathbf {E})\{(\, \mathbf
{i}_\alpha  \mathbf {H}) \mathbf {H}- \mathbf {i}_\alpha  \mathbf
{H}^2\}.\tag{A3}
\end{equation*}
As one can see, the vector $[\mathbf {u}_1^{(\alpha )} \times
\mathbf {H}]$ lies in the plane specified by the vectors $\mathbf
{H}$ and $\mathbf {i}_\alpha$. If we take the direction $[\mathbf
{i}_\alpha \times \mathbf {H}]$ in this plane as the $x$ axis and
the direction $\mathbf {i}_\alpha $ as the $z$ axis (the normal to
these vectors will be the $y$ axis), then the components $[\mathbf
{u}_1^{(\alpha )} \times \mathbf {H}]_y $ and $[\mathbf
{u}_1^{(\alpha )} \times \mathbf {H}]_z$ will be non-zero. Their
substitution into Eqs. (\ref{eq3}) and (\ref{eq4}), respectively,
yields
\begin{equation*}
\mathbf {u}_{2y}^{(\alpha )} =\frac{e^3\tau _ \perp ^2 }{m_ \perp ^2
c^2}(\tau _{\parallel} /m_{\parallel} -\tau _ \perp /m_ \perp )(
\mathbf {i}_\alpha  \mathbf {E})( \mathbf {i}_\alpha  \mathbf {H})\{
\mathbf {H}- \mathbf {i}_\alpha ( \mathbf {i}_\alpha  \mathbf
{H})\},\tag{A4}
\end{equation*}
\begin{equation*}
\mathbf {u}_{2z}^{(\alpha )} \!=-\frac{e^3\tau _ \perp \tau
_{\parallel} }{m_ \perp m_{\parallel} c^2}(\tau _{\parallel}
/m_{\parallel} -\tau _ \perp /m_ \perp )( \mathbf {i}_\alpha \mathbf
{E})\{H^2-( \mathbf {i}_\alpha \mathbf {H})^2\} \mathbf {i}_\alpha
.\tag{A5}
\end{equation*}
Expressions (А4) and (А5) are put down in the vector form so that to
be easy-to-use in the laboratory system of coordinates and to sum up
over all valleys. Thus, $\mathbf {u}_{2y}^{(\alpha )}$ and $\mathbf
{u}_{2z}^{(\alpha )} $ are related to the given ellipsoid $\alpha $
only through the unit vector $\mathbf {i}_\alpha $.

As we are interested in the longitudinal magnetoresistance, it is
necessary to find the addition to the current $J_0 $ quadratic in
the magnetic field (in the direction $\mathbf {E}\parallel\mathbf
{H}\mathbf {q}_0; \mathbf {q}_0 \equiv $ (1,0,0)). This addition is
evidently equal to
\begin{equation*}
\Delta J_2 =- \frac{en}{4}\sum\limits_{(\alpha )} { \mathbf {q}_0
\{\mathbf {u}_{2y}^{(\alpha )} + \mathbf {u}_{2z}^{(\alpha )}
\}}.\tag{A6}
\end{equation*}
Here, $\frac{1}{4}n$ is the electron concentration in one valley.

Substituting expressions (А4) and (А5) into (А6) and taking into
account that the unit vectors $\mathbf {i}_\alpha  (\alpha
=1,2,3,4)$ in $n$-Ge have the form
\[
 \mathbf {i}_1 =\frac{1}{\sqrt 3 }\,(1,1,1),
\quad
 \mathbf {i}_2 =\frac{1}{\sqrt 3 }\,(-1,1,1),
\]
\[
 \mathbf {i}_3 =\frac{1}{\sqrt 3 }\,(1,-1,1),
\quad
 \mathbf {i}_4 =\frac{1}{\sqrt 3 }\,(-1,-1,1),
\]
we obtain
\begin{equation*}
\Delta  \mathbf {J}_2 = -\frac{e^4n}{9m_ \perp \,c^2}\left(
{\frac{\tau _{\parallel} }{m\parallel}-\frac{\tau _ \perp }{m_ \perp
}} \right)^2H^2 \mathbf {E}.\tag{A7}
\end{equation*}
For isotropic scattering mechanisms and dispersion law, $\Delta
J_2 $ = 0.

}

\rezume{%
ВПЛИВ СЛАБКОГО МАГНІТНОГО\\ ПОЛЯ ($\sim $300 Гс) НА
ІНТЕНСИВНІСТЬ\\ ТЕРАГЕРЦОВОГО ВИПРОМІНЮВАННЯ ГАРЯЧИХ\\
ЕЛЕКТРОНІВ В $n$-Ge  ПРИ ГЕЛІЄВИХ ТЕМПЕРАТУРАХ}{В.М. Бондар, П.М.
Томчук, Г.А. Шепельський} {У роботі наведено експериментальні
результати та їх обговорення у вивченні впливу слабкого магнітного
поля ($\sim $300  Гс) на інтенсивність терагерцового випромінювання
($\lambda \approx $100 мкм) гарячих електронів з $n$-Ge
(кристалографічний напрямок ($\langle 1,0,0 \rangle)$\rule{0pt}{9pt}
при гелієвих температурах\rule{0pt}{9pt} $T\sim $5 К). Показано, що
сильний вплив такого поля (зменшення\rule{0pt}{9pt} інтенсивності)
випромінювання (500--1000 {\%}) пов'язаний зі
зменшенням\rule{0pt}{9pt} концентрації носіїв при\rule{0pt}{9pt}
слабких електричних полях та появою магнітоопору при сильніших
полях.\rule{0pt}{9pt} Поздовжній магнітоопір стає\rule{0pt}{9pt}
суттєвим\rule{0pt}{9pt} завдяки анізотропії закону дисперсії енергії
електронів\rule{0pt}{9pt} і сильній деформації електричним полем
функції розподілу електронів\rule{0pt}{9pt} за швидкостями (вихід за
межі дифузійного наближення).\rule{0pt}{9pt}}

\end{document}